\documentclass[reqno]{amsart}
\usepackage[latin1]{inputenc}
\usepackage{amsthm}
\usepackage{amssymb}
\usepackage{latexsym}
\usepackage{pictex}
\usepackage[mathscr]{eucal}
\newtheorem{theorem}{Theorem}[section]

\newtheorem{corollary}[theorem]{Corollary}

\newtheorem{definition}[theorem]{Definition}
\newcommand{\dist}{\mbox{\rm dist}}
\newcommand{\supp}{\mbox{\rm supp} \,}
\renewcommand{\epsilon}{\varepsilon}
\newcommand{\PP}{\mathbb{P}}
\newcommand{\EE}{\mathbb{E}}
\newcommand{\RR}{\mathbb{R}}
\newcommand{\ZZ}{\mathbb{Z}}
\newcommand{\CC}{\mathbb{C}}
\newcommand{\NN}{\mathbb{N}}
\renewcommand{\aa}{{\bf a}}
\newcommand{\ao}{{\bf a}_\omega}
\newcommand{\diag}{\mbox{\rm diag}}

\begin{document}
\title{Multi-scale analysis implies strong dynamical localization}
\author[D.~Damanik and P.~Stollmann]{David Damanik$\,^{1,2}$ and Peter Stollmann$\,^{1}$}
\maketitle
\vspace{0.3cm}    
\noindent    
$^1$ Fachbereich Mathematik, Johann Wolfgang Goethe-Universit\"at,    
60054 Frankfurt, Germany\\[0.2cm]    
$^2$ Department of Mathematics 253--37, California Institute of Technology,    
Pasadena, CA 91125, U.S.A.\\[0.3cm]    
E-mail: damanik@its.caltech.edu, stollman@math.uni-frankfurt.de
\setcounter{section}{0}
\renewcommand{\theequation}{\arabic{section}.\arabic{equation}}
\begin{abstract}
We prove that a strong form of dynamical localization follows from a variable energy multi-scale analysis. This abstract result is applied to a number of models for wave propagation in disordered media.
\end{abstract}

\section{Introduction}

In the present paper we prove that a variable energy multi-scale
analysis implies dynamical localization in a strong (expectation)
form. Thus we accomplish a goal of a long line of research. Ever since
Anderson's paper \cite{anderson}, the dynamics of waves in random media
has been a subject of intensive research in mathematical physics. The
breakthrough as far as mathematically rigorous results are concerned came 
with the paper \cite{fs} by Fröhlich and Spencer in which absence of
diffusion is proven. They also introduced a technique of central
importance to the topic: multi-scale analysis.

The next step was a proof of exponential localization, by which one
understands pure point spectrum with exponentially decaying
eigenfunctions; see the bibliography for a list of results in different
generality. However, from the point of view of transport properties,
exponential localization does not yield too much information. We refer to
\cite{drjls,drjls1} where a strengthening of exponential decay is
introduced, property (SULE), which in fact allows one to prove dynamical 
localization. In \cite{gdb} it was shown that a variable energy
multi-scale analysis implies (SULE) almost surely, so that
$$
\sup_{t>0}\| |X|^pe^{-itH(\omega)}P_I(H(\omega)) \chi_K \| < \infty \quad
\mbox{$\PP$-a.s.},
$$
where $H(\omega)$ is a random Hamiltonian which admits multi-scale
analysis in the interval $I$, $P_I$ is the spectral projector onto that
interval, and $K$ is compact.

We will strengthen the last statement to 
$$
\EE\Big\{\sup_{t>0}\|
|X|^pe^{-itH(\omega)}P_I(H(\omega))\chi_K \| \Big\} < \infty.
$$
Here, as in \cite{gdb}, the $p$ which is admissible depends on the
characteristic parameters of multi-scale analysis. In order to explain
this we will sketch in the next section an abstract form of multi-scale 
analysis and introduce the necessary set-up. In Section 3, we show that multi-scale analysis implies dynamical localization in the expectation. We do so by showing that (more or less) for $\eta \in L^\infty$, $\supp \eta \subset I$ (the localized region),
$$
\EE\Big\{\|\chi_{\Lambda_1}\eta(H(\omega))\chi_{\Lambda_2}\|\Big\}\le\|\eta\|_{\infty} \cdot \dist(\Lambda_1,\Lambda_2)^{-2\xi},
$$
where $\xi$ is one of the characteristic exponents of multi-scale
analysis. We should note here that the main progress concerns continuum models, since for discrete models the Aizenman technique \cite{a,am} is available, which gives even exponential decay of the expectation above (see \cite{ag} for an exposition in which a number of applications is presented and the very recent \cite{asfh} which shows that the Aizenman technique is
applicable in the energy region in which multi-scale analysis works). However, our results clearly apply to discrete models with singular single-site distribution, most notably the one-dimensional Bernoulli-Anderson model. Moreover, we refer to \cite{bfm} where a study of time means instead of the sup is undertaken. However, the latter paper does not contain too much about continuum models, and the results we present contain the estimates given there. In Section 4 we present our applications to a number of models for wave propagation in disordered media, including band edge dynamical localization for Schr\"{o}dinger and divergence form operators as well as Landau Hamiltonians.

\section{The multi-scale scenario}
In this section we present the abstract framework for multi-scale
analysis developed in \cite{book1}. We start with a number of properties 
which are easily verified for the applications we shall discuss, where $H(\omega)$ is a random operator in $L^2(\RR^d)$ and $H_\Lambda(\omega)$ denotes its restriction to an open cube $\Lambda\subset\RR^d$ with suitable boundary conditions. 

We call a cube $\Lambda = \Lambda_L(x)$ of sidelength $L$ centered at $x$
{\em suitable} if $x\in\ZZ^d$ and $L \in 3\NN \setminus 6\NN$. In this case
$\overline{\Lambda}$ itself as well as $\overline{\Lambda}_{L/3}(x)$ are unions of closed unit cubes centered on the lattice. Denote 
$$
\Lambda^{{\rm int}}:=\Lambda_{L/3}(x), \;
\Lambda^{{\rm out}}:=\Lambda_L(x)\setminus\Lambda_{L-2}(x),
$$
and denote the respective characteristic functions by $\chi^{{\rm int}}=\chi_\Lambda^{{\rm int}} = \chi_{L,x}^{{\rm int}} := \chi_{\Lambda^{{\rm int}}}$ and $\chi^{{\rm out}}=\chi_\Lambda^{{\rm out}} = \chi_{L,x}^{{\rm out}} := \chi_{\Lambda^{{\rm out}}}$.

The first condition concerns measurability and independence:
\\[5mm]
{\bf (INDY)} $(\Omega,\mathcal{F},\PP)$ is a probability space; for every cube $\Lambda$, $H_\Lambda(\omega)$ is a self-adjoint operator in $L^2(\Lambda)$, measurable in $\omega$, such that $H_{\Lambda_L(x)}(\omega)$ is stationary in $x\in\ZZ^d$ and $H_\Lambda$ and $H_{\Lambda'}$ are independent for disjoint cubes $\Lambda$ and $\Lambda'$.\\[3mm]
\indent
So far, $H_\Lambda$ and $H_{\Lambda'}$ are not related if $\Lambda \subset \Lambda'$. The next condition supplies a relation. In concrete examples it is the so-called geometric resolvent inequality which follows from commutator estimates and the resolvent identity. For $E \in \rho (H_\Lambda (\omega))$, we denote
$$
R_\Lambda(E) = R_\Lambda(\omega,E) = (H_\Lambda(\omega)-E)^{-1}.
$$
{\bf (GRI)} For given bounded $I_0 \subset \RR$, there is a constant $C_{{\rm geom}}$ such that for all suitable cubes $\Lambda,\Lambda'$ with $\Lambda \subset \Lambda'$, $A \subset \Lambda^{{\rm int}}$, $B \subset \Lambda' \setminus \Lambda$, $E \in I_0$ and $\omega \in \Omega$, the following inequality holds:
$$
\|\chi_B R_{\Lambda'}(E) \chi_A \| \le C_{{\rm geom}} \cdot \| \chi_B R_{\Lambda'}(E) \chi_\Lambda^{{\rm out}} \| \cdot \| \chi_\Lambda^{{\rm out}} R_\Lambda(E)\chi_A\|.
$$

Finally, we need an upper bound for the trace of the local Hamiltonians
$H_\Lambda$ in a given bounded energy region $I_0$, which follows from
Weyl's law in concrete cases at hand.
\\[3mm]
{\bf (WEYL)} For each interval $J\subset I_0$, there is a constant $C$ such that 
$$
\mbox{tr} (P_J(H_\Lambda(\omega))\le C\cdot|\Lambda|\ \mbox{ for all }
\omega\in\Omega.
$$

Here $P_J(\cdot)$ denotes the spectral projection of the operator in
question. Given this basic set-up, multi-scale analysis deals with an
inductive proof of resolvent decay estimates. This resolvent decay is
measured in terms of the following concept:

\begin{definition}
Let $\Lambda=\Lambda_L(x)$, $x\in\ZZ^d$, $L \in 2\NN+1$. $\Lambda$ is called $(\gamma,E)$-{\em good for\/} $\omega\in\Omega$ if 
$$
\|\chi^{{\rm out}} R_\Lambda(E)\chi^{{\rm int}}\|\le\exp(-\gamma\cdot L).
$$
$\Lambda$ is called $(\gamma,E)$-{\em bad for\/} $\omega\in\Omega$ if it is not $(\gamma,E)$-good for $\omega$.
\end{definition}

We can now define the property on which we base our induction:
\\[3mm]
{\bf\boldmath $G(I,L,\gamma,\xi)$} $\forall x,y\in\ZZ^d$,
$d(x,y)\ge L$ the following estimate holds:
\[
\PP\{\forall E\in I:\Lambda_L(x)\ \mbox{ or } \Lambda_L(y)\ \mbox{ is }
(\gamma,E)\mbox{-good for } \omega\}\ge1-L^{-2\xi}.
\]

The basic idea of the multi-scale induction is that we consider some
larger cube $\Lambda'$ with sidelength $L'=L^\alpha$. With high
probability there are not too many disjoint bad cubes of sidelength $L$
in $\Lambda'$. Of course, since the number of cubes in $\Lambda'$ is
governed by $\alpha$, this will only hold if $\alpha$ is not too large, 
depending on $\xi$.

By virtue of the geometric resolvent inequality (GRI), each of the good
cubes of sidelength $L$ in $\Lambda'$ will add to exponential decay on the 
big cube. In order to make this work, we will additionally need a 
``worst case estimate.'' This is given by the following weak form of a
Wegner estimate: 
\\[3mm]
{\bf\boldmath $W(I,L,\Theta,q)$} For all $E\in I$ and
$\Lambda=\Lambda_L(x)$, $x\in\ZZ^d$, the following estimate holds:
\[
\PP\{\dist(\sigma(H_\Lambda(\omega)),E)\le\exp(-L^\Theta)\}\le L^{-q}.
\]

We have the following theorem:

\begin{theorem}\label{t21}
Let $I_0 \subset \RR$ be a bounded open set and assume that $H_\Lambda(\omega)$ satisfies {\rm (INDY), (GRI)} and {\rm (WEYL)} for $I_0$.

Assume that there are $L_0 \in 2 \NN + 1$, $q>d$, $\Theta \in (0,\frac12)$
such that for $L\ge L_0$, $L \in 2\NN + 1$, the Wegner estimate
$W(I_0,L,\Theta,q)$ is valid.

Furthermore, fix $\xi_0>0$ and $\beta>2\Theta$. Let $\alpha\in(1,2)$ be such that
$$
4d \, \frac{\alpha-1}{2-\alpha} \le \xi_0 \wedge \frac14(q-d) .
$$
Then there exist $C_1=C_1(d,C_{{\rm geom}})$ and $L^* = L^*(q, d, \xi_0, \Theta, \beta, \alpha)$ such that the following implication
holds:

If for $\overline I \subset I_0$, $L \ge L^*$, $L \in 3 \NN \setminus 6 \NN$, and $\gamma_L \ge L^{\beta-1}$, the estimate $G(I,L,\gamma_L,\xi_0)$ is satisfied, then $G(I,L',\gamma_{L'},\xi)$ also holds, where
\begin{enumerate}
\item[{\rm (i)}] $L' \in 3 \NN \setminus 6 \NN$, $L^\alpha \le L' \le L^\alpha+6$,
\item[{\rm (ii)}] $\xi \ge \xi_0 \wedge [\frac14(q-d)]$,
\item[{\rm (iii)}] $\gamma_{L'} \ge \gamma_L (1-8 L^{1-\alpha}) - C_1 \cdot
L^{-1} - 6 L^{\alpha(\Theta-1)} \ge (L')^{1-\beta}$.
\end{enumerate}
\end{theorem}

For a proof of the result in this form we refer to \cite{book1}. It is modelled after the variable multi-scale analysis by von Dreifus-Klein \cite{vdk}. See also \cite{fk} and \cite{gdb} for continuum versions.

Let us now formulate an immediate consequence of the preceding theorem.

\begin{corollary}\label{c22}
Let $I_0, (H_\Lambda(\omega)), \xi_0, \beta, q, \Theta, \alpha \in (1,2)$ be as in Theorem {\rm \ref{t21}}. There exists $\overline L = \overline
L(\xi_0, \beta, \Theta, q, C_{{\rm geom}}, \alpha)$ such that the following holds. 

If $I \subset I_0$ and $G(I,L,\gamma_L,\xi_0)$ is satisfied for some
$\gamma_L \ge L^{\beta-1}$ and some $L \ge \overline L$, then there exist a
sequence $(L_k)_{k\in \NN} \subset 3 \NN \setminus 6 \NN$ and $\gamma_\infty > 0$ with the following properties:
\begin{enumerate}
\item[{\rm (i)}] For all $k \in \NN$, the estimate $G(I,L_k,\gamma_\infty,\xi)$ is satisfied, where $\xi = \xi_0 \wedge \frac14(q-d)$.
\item[{\rm (ii)}] $L_k^\alpha \le L_{k+1} \le L_k^\alpha+6$.
\end{enumerate}
\end{corollary}

\section{Multi-scale estimates imply strong dynamical localization}
We keep the framework introduced in the preceding section. Thus we start 
out with a family $H_\Lambda$ of random local Hamiltonians where
$\Lambda$ runs through the suitable cubes. Now we introduce a link to a
Hamiltonian on the whole space $\RR^d$. Consider the statement
\\[5mm]
{\bf (EDI)} Assume that $H(\omega)$ is a self-adjoint operator in
$L^2(\RR^d)$, measurable with respect to $\omega$, and suppose that there is a measurable set $\Omega_1$ with $\PP(\Omega_1)=1$ and a constant $C_{{\rm EDI}}$ such that for every $\omega\in\Omega_1$, the spectrum of $H(\omega)$ in $I_0$ is pure point and every eigenfunction $u$ of $H(\omega)$ corresponding to $E\in I_0$ satisfies
$$
\|\chi_\Lambda^{{\rm int}}u\|\le C_{{\rm EDI}} \cdot \|\chi_\Lambda^{{\rm out}} (H_\Lambda(\omega)-E)^{-1} \chi_\Lambda^{{\rm int}}u\| \cdot \|\chi_\Lambda^{{\rm out}} u\|. \eqno{{\rm (EDI)}} 
$$

For the operators $H(\omega)$ we shall consider in Section 4 and $H_\Lambda (\omega)$ the restriction to $\Lambda$ with respect to suitable boundary conditions, the eigenfunction decay inequality (EDI) readily follows. Moreover, in this case, we can use the multi-scale machinery to prove pure point spectrum almost surely. Therefore, the condition above seems to be a natural abstract condition. We can now state the main result of the present paper:

\begin{theorem}\label{t31}
Assume that $H(\omega)$ and $H_\Lambda(\omega)$ satisfy {\rm (INDY), (GRI), (WEYL)} and {\rm (EDI)} above for a given bounded open set $I_0\subset\RR$. Moreover, assume

\begin{itemize}
\item[{\rm (i)}] $\chi_\Lambda P_{I_0}(H(\omega))\chi_\Lambda$ is trace class for every suitable cube $\Lambda$ and 
$$
\mbox{{\rm tr}} (\chi_\Lambda P_{I_0}(H(\omega))\le C_{{\rm tr}} \cdot | \Lambda |^{\kappa}
$$
for some fixed $\kappa$.
\item[{\rm (ii)}] There exist $L_0\in\NN$, $q>d$ and $\Theta\in(0,\frac12)$
such that for $L\in3\NN\setminus 6\NN$, $L\ge L_0$, the Wegner estimate 
$W(I_0,L,\Theta,q)$ is valid.
\item[{\rm (iii)}] For $q>d$ from the Wegner estimate and $\xi_0>0$  we have that
$$
p<2\xi_0\wedge\tfrac14(q-d).
$$
\end{itemize}
Then there exists $\overline{L}= L(p,\xi_0, \beta, \Theta, q, C_{{\rm
    geom}}, d)$ such that if
\begin{itemize}
\item[{\rm (iv)}] for some $L \in 3 \NN \setminus 6\NN$, $L \ge
  \overline L$, there is an open interval $I \not= \emptyset$, $I
  \subset I_0$, such that $G(I,L,L^{\beta-1},\xi_0)$ holds,
\end{itemize}
then for every $\eta \in L^\infty$ with $\supp \eta \subset I$, it follows that
$$
\EE\{\| |X|^p \eta(H(\omega))\chi_K\|\}<\infty
$$
for every compact set $K\subset\RR^d$.
\end{theorem}

Let us first sketch the idea of the proof which is quite simple. Of
course, by $|X|^p$ we denote the operator of multiplication with $|x|^p$.

We write
\begin{equation}\label{eq132}
\| \chi_{\Lambda_1} \eta(H(\omega)) \chi_{\Lambda_2} \| \le \sum_{E_n \in I_0} \| \chi_{\Lambda_1} \phi_n(\omega) \| \cdot \| \chi_{\Lambda_2} \phi_n(\omega) \| \cdot | \eta(E_n(\omega)) |,
\end{equation}
where $E_n(\omega)$, $\phi_n(\omega)$ denote the eigenvalues and
eigenfunctions of $H(\omega)$ in $I$. The probability that both
$\Lambda_1$ and $\Lambda_2$ are bad for the same $E_n(\omega)$ is small, 
roughly polynomially in the distance between $\Lambda_1,\Lambda_2$. If
one of them is good, the eigenfunction decay inequality (EDI) says that one of the norms appearing in the rhs of (\ref{eq132}) is exponentially
small. This leads to a polynomial decay of $\EE\{\|\chi_{\Lambda_1}\eta(H(\omega))\chi_{\Lambda_2}\|\}$ once the
interval $I$ is suitably chosen to guarantee the necessary probabilistic 
estimates. The assumption in (iii) of the theorem ensures that the
polynomial growth of $|X|^p$ is killed by this polynomial decay. To make 
all of this work we have to overcome the difficulty that in the sum in
(\ref{eq132}) we have infinitely many terms. This is taken care of by
analyzing the centers of localization $x_n(\omega)$ of $\phi_n(\omega)$. All this will be done relatively to a certain length scale $L_k$.

We proceed in several steps. The first steps will be used to choose an
appropriate $\alpha$ and set up a multi-scale scenario. Then we take
care of those $\phi_n(\omega)$ whose centers are far away from $K$. To
this end, we employ the Weyl-type trace condition (i).

\begin{proof}[{\bf Proof.}]
{\bf Step 1.} Choose $\alpha\in(1,2)$ such that
$$
4d\, \frac{\alpha-1}{2-\alpha}\le\frac14(q-d)\wedge\xi_0=:\xi
$$
and
$$
3d(\alpha-1)+\alpha p<2\xi.
$$
Note that the latter condition can be achieved for $\alpha > 1$ small
enough, since $p < 2 \xi$. For this choice of $\alpha$, let
$\overline{L}$ be the minimal length scale from Corollary \ref{c22}.

We can now use Theorem \ref{t21} and Corollary \ref{c22} to find a sequence $(L_k)_{k \in \NN} \subset \NN$ and a constant $\gamma > 0$ such that for every $k$, 

\begin{itemize}
\item $L_k \in 3 \NN \setminus 6 \NN$,
\item $L^\alpha_k \le L_{k+1} \le 6 L_k^\alpha$,
\item $G(I,L_k,\gamma,\xi)$ is satisfied.
\end{itemize}
For $j\in\NN$, denote $\Gamma_j=\left( \frac{L_j}3 \ZZ \right)^d$ and 
\begin{align*}
E_j  = & \, \{\omega \in \Omega;\ \mbox{for some } E\in I\ \mbox{there exist } y,z \in \Gamma_j \cap \Lambda_{3L_{j+1}} \mbox{ such that } \Lambda_{L_j}(y)\\ & \mbox{ and } \Lambda_{L_j}(z) \mbox{ are disjoint and both not } \mbox{$(\gamma,E)$-good} \}.
\end{align*}
Since $\Gamma_j \cap \Lambda_{3 L_{j+1}} \le \Big( \frac{9L_{j+1}}{L_j} \Big)^d \le (54)^d L_j^{d(\alpha-1)}$ and $G(I,L_j,\gamma,\xi)$ holds, we have
$$
\PP(E_j)\le c_dL_j^{2d(\alpha-1)-2\xi}.
$$
For $k\in\NN$, denote
$$
\Omega^k_{{\rm 2bad}}=\bigcup_{j\ge k}E_j.
$$

{\bf Claim.} For every $k \in \NN$,

\begin{equation}\label{eq133}
\PP(\Omega^k_{{\rm 2bad}})\le c(\alpha,d,\xi)\cdot L_k^{2d(\alpha-1)-2\xi}.
\end{equation}

{\it Proof.} We have

\begin{align*}
\PP(\Omega^k_{{\rm 2bad}}) \le& \, c_d \cdot \sum_{j\ge k} L_j^{2d(\alpha-1)-2\xi}\\
\le & \, c_d \cdot L_k^{2d(\alpha-1)-2\xi} \cdot \Big( 1+ \sum_{j\ge k+1} \Big( \frac{L_j}{L_k} \Big)^{2d(\alpha-1)-2\xi}\Big).  
\end{align*}
Now, for $j\ge k+1$,
$$
\frac{L_j}{L_k}\ge\frac{L_k^{\alpha^{j-k}}}{L_k}=L_k^{\alpha^{j-k}-1}\ge3^{\alpha^{j-k}},
$$
which gives the assertion. 

{\bf Step 2.} Denote by $\phi_n(\omega)$ the normalized
eigenfunctions of $H(\omega)$, $\omega\in\Omega_1$ with corresponding
eigenvalues $E_n(\omega)\in I$. For each $\omega,n$, define a {\em center
of localization\/} $x_n(\omega)\in\ZZ^d$ by
$$\| \chi_{\Lambda_1(x_n(\omega))} \phi_n(\omega) \| = \max \{ \| \chi_{\Lambda_1(y)} \phi_n(\omega) \| ; y \in \ZZ^d \}.
$$
Since $\phi_n(\omega)\in L^2$, such a center always exists.

{\bf Claim.} There is $k_0=k_0(\gamma,d,C_{{\rm EDI}})$ such that for
$\omega\in\Omega_1$, $k\ge k_0$ and $x_n(\omega)\in\Lambda_{L_k}^{{\rm int}}(x)$, the cube $\Lambda_{L_k}(x)$ is $(\gamma,E_n(\omega))$-bad.

{\it Proof.} Assume otherwise. Then by (EDI) it follows that
$$
\|\chi_{\Lambda_1(x_n(\omega))} \phi_n(\omega) \| \le \| \chi^{{\rm int}}_{L_k,x} \phi_n(\omega) \| \le C_{{\rm EDI}} \cdot e^{-\gamma L_k} \cdot \| \chi^{{\rm out}}_{L_k,x} \phi_n(\omega) \|.
$$
Estimating the number of unit cubes in $\Lambda^{{\rm out}}_{L_k}(x)$ very
roughly by $L_k^d$ we find that
$$
\cdots \le C_{{\rm EDI}}\cdot e^{-\gamma L_k} \cdot L_k^d \cdot \max_{\tilde x \in \Lambda^{{\rm out}}_{L_k}(x)} \| \chi_{\Lambda_1(\tilde x)} \phi_n(\omega)\|.
$$
If $k_0$ is large enough to ensure 
$$
C_{{\rm EDI}} \cdot e^{-\gamma L_{k_0}} \cdot L_{k_0}^d<1,
$$
the inequality above contradicts the choice of $x_n(\omega)$.

{\bf Step 3:} Let $\omega \in \Omega^k_{{\rm 2good}} = (\Omega^k_{{\rm 2bad}})^c \cap \Omega_1$ with $k\ge k_0$. Then there exists
$j_0=j_0(\gamma,\alpha,d, C_{{\rm EDI}})$ such that for $j\ge j_0$, $j\ge k$ and $x_n(\omega) \in \Lambda_{L_{j+1}}$,
$$
\|(1-\chi_{3L_{j+2}})\phi_n(\omega)\|^2 \le\tfrac14,
$$
where $\chi_L$ is shorthand for $\chi_{\Lambda_L(0)}$.

{\it Proof.} We divide $\Lambda^c_{3L_{j+2}}$ into angular regions $M_i$, 
$$
M_i=\Lambda_{3L_{i+1}}\setminus \overline\Lambda_{3L_i},\ i\ge j+2.
$$
We have 
\begin{align*}
\|(1-\chi_{3L_{j+2}})\phi_n(\omega)\|^2=&\sum_{i\ge j+2} \|\chi_{M_i}\phi_n(\omega)\|^2\\
=&\sum_{i\ge j+2} \, \sum_{\tilde x \in M_i\cap\Gamma_i} \| \chi^{{\rm int}}_{L_i,\tilde x}\phi_n(\omega)\|^2.
\end{align*}
By construction of $M_i$, for every $\tilde x\in M_i\cap\Gamma_i$, we
find $\tilde x_n\in\Gamma_i\cap\Lambda_{L_{j+1}}$ such that
$x_n(\omega)\in\Lambda^{{\rm int}}_{L_i}(\tilde x_n)$ and $d(\tilde x, \tilde
x_n)\ge L_i$.

Since $\Lambda_{L_i}(\tilde x_n)$ is $(\gamma,E_n(\omega))$-bad and
$\omega\in\Omega^k_{{\rm 2good}}$, it follows that $\Lambda_{L_i}(\tilde x)$
is $(\gamma, E_n(\omega))$-good so that 
$$
\|\chi^{{\rm int}}_{L_i,\tilde x}\phi_n\|^2 \le (C_{{\rm EDI}})^2 \cdot e^{-2\gamma L_i}.
$$
Since $\# M_i\cap\Gamma_i$ grows only polynomially in $L_i$, the assertion follows.

{\bf Step 4:} There exists $C = C(\gamma,\alpha,d,\kappa,C_{{\rm tr}})$
such that for $\omega\in\Omega^k_{{\rm 2good}}$, $j\ge k$,
$$
\#\{n; x_n(\omega)\in\Lambda_{L_{j+1}}\}\le C\cdot L_{j+1}^{\alpha\kappa
    d}.
$$

{\it Proof.} Since $\#\{\ldots\}$ is non-decreasing in $j$, and since $j_0$
from Step 3 only depends on $(\gamma, \alpha,d)$, we can restrict
ourselves to the case $j\ge j_0$ and adapt the constant $C$.

We start by observing
$$
\sum_{x_n\in\Lambda_{L_{j+1}}}(\chi_{3 L_{j+2}} P_I(H(\omega)) \chi_{3 L_{j+2}} \phi_n(\omega)| \phi_n(\omega)) \le \mbox{tr}( \chi_{3 L_{j+2}} P_I(H(\omega))).
$$
We want to show that each of the terms in the sum is at least $\frac12$,
thus giving an estimate on the number as asserted. Using Step 3 and suppressing $\omega$, we have
\begin{align*}
(\chi_{3 L_{j+2}}P_I\chi_{3 L_{j+2}}\phi_n|\phi_n)=&\, (\chi_{3 L_{j+2}}P_I\phi_n|\phi_n)-(\chi_{3 L_{j+2}}P_I(1-\chi_{3 L_{j+2}})\phi_n|\phi_n)\\
\ge& \, (\chi_{3 L_{j+2}}\phi_n|\phi_n)-\tfrac14\\
=&\, (\phi_n|\phi_n)-((1-\chi_{3 L_{j+2}})\phi_n|\phi_n)-\tfrac14\\
\ge&\, \tfrac12.
\end{align*}
Plugging this into the above estimate on the trace, we get the claimed
bound for the number ${\#}\{n;\ldots\}$.

{\bf Step 5:} There is $k_1=k_1(C_{{\rm EDI}}, \alpha, C_{{\rm tr}}, \kappa, L_0, \gamma, d)$ such that for $k \ge k_1$, $\omega \in \Omega^k_{{\rm 2good}}$ and $x \in \Gamma_k \cap \Lambda_{L_{k+1}} \setminus \Lambda_{L_k}$,
$$
\|\chi^{{\rm int}}_{L_k,x}\eta (H(\omega))\chi^{{\rm int}}_{L_k,0}\|\le
\exp(-\tfrac\gamma2 L_k)\cdot\|\eta\|_\infty.
$$

{\it Proof.} We have

\begin{equation}\label{eq134}
\|\chi^{{\rm int}}_{L_k,x}\eta(H(\omega))\chi^{{\rm int}}_{L_k,0} \| \le \sum_{E_n\in I}| \eta(E_n(\omega))| \cdot \|\chi^{{\rm int}}_{L_k,x} \phi_n(\omega) \| \cdot \|\chi^{{\rm int}}_{L_k,0}\phi_n(\omega)\|.
\end{equation}
We now divide the sum according to where the $x_n(\omega)$ are located:
$$
\sum_{\substack{E_n\in I \\ x_n(\omega) \in \Lambda_{k+1}}} \|\chi^{{\rm int}}_{L_k,x} \phi_n(\omega) \| \cdot \| \chi_{L_k,0}^{{\rm int}} \phi_n(\omega) \| \le C \cdot L_{k+1}^{\alpha\kappa d} \cdot C_{{\rm EDI}} \cdot e^{-\gamma L_k},
$$
since one of the cubes $\Lambda_{L_k}(x),\Lambda_{L_k}(0)$ has to be
$(\gamma,E_n(\omega))$-good and the number of $x_n(\omega)$ has been estimated in Step 4.

For $k$ large enough, depending only on the indicated parameters, $k\ge k_0$,
\begin{equation}\label{eq135}
\sum_{ \substack{E_n\in I \\ x_n(\omega) \in \Lambda_{k+1} } } \|\chi^{{\rm int}}_{L_k,x}\phi_n(\omega)\|\cdot\|\chi_{L_{k,0}}^{{\rm int}}\phi_n(\omega)\| \le \tfrac12\exp(-\tfrac\gamma2L_k). 
\end{equation}
We now treat the remaining terms. Note that for $j\ge k+1$ and
$x_n(\omega)\in\Lambda_{L_{j+1}}\setminus\Lambda_{L_j}$, we find an
$\tilde x_n(\omega)\in\Lambda_{L_{j+1}}\cap\Gamma_j$ such that
$x_n(\omega)\in\Lambda_{L_j}^{{\rm int}}(\tilde x_n(\omega))$. From Step 2 we know that $\Lambda_{L_j}(\tilde x_n(\omega))$ must be $(\gamma,E_n(\omega))$-bad so that $\Lambda_{L_j}(0)$ has to be $(\gamma, E_n(\omega))$-good since $\omega\in\Omega^k_{{\rm 2good}}$. Therefore
$$
\| \chi^{{\rm int}}_{L_k,0} \phi_n(\omega) \| \le \| \chi^{{\rm int}}_{L_j,0} \phi_n(\omega) \| \le C_{{\rm EDI}} \cdot \exp 
(-\gamma L_j).
$$
Using Step 4 again, we see that

\begin{align*}
\sum^{\infty}_{j=k+1} \Big( \sum_{x_n\in\Lambda_{L_{j+1}} \setminus \Lambda_{L_j}} \| \chi^{{\rm int}}_{L_k,x} \phi_n(\omega) \| \cdot \|\chi^{{\rm int}}_{L_k,0} \phi_n(\omega) \| \Big) \le & \, C \cdot C_{{\rm EDI}} \cdot \sum^\infty_{j=k+1} e^{-\gamma L_j} L_{j+1}^{\alpha\kappa d}\\
\le &\, \tfrac12 \exp(-\tfrac\gamma2 L_k)
\end{align*}
if $k \ge k_1(C_{{\rm EDI}}, \alpha, C_{\rm tr}, \kappa, L_0, \gamma, d)$. The latter estimate, together with (\ref{eq134}) and (\ref{eq135}), gives the assertion.

{\bf Step 6:} For $k\ge k_1$ from Step 5 and $x \in \Gamma_k \cap \Lambda_{L_{k+1}} \setminus \Lambda_{L_k}$, we have
$$
\EE \{ \| \chi^{{\rm int}}_{L_k,x} \eta(H(\omega)) \chi^{{\rm int}}_{L_k,0} \| \} \le \| \eta \|_\infty \cdot \Big (c(\alpha, d, \xi) \cdot L_k^{2d (\alpha-1) - 2 \xi} + \exp(-\tfrac\gamma2 L_k)\Big) .
$$

{\it Proof.} For $\omega \in \Omega^k_{{\rm 2bad}}$, we can estimate the norm by $\|\eta\|_\infty$ and use Step 1, while for $\omega\in\Omega^k_{{\rm 2good}}$, we can use Step 5.

Put together, we have
\begin{align*}
\EE \{ \ldots \} \le &\, \|\eta\|_\infty \cdot \Big( \PP( \Omega^k_{{\rm 2bad}}) + \exp(-\tfrac\gamma2L_k) \PP( \Omega^k_{{\rm 2good}}) \Big)\\
\le &\, \|\eta\|_\infty \cdot \Big(c(\alpha,d,\xi)L_k^{2d (\alpha-1)- 2 \xi}+ \exp(-\tfrac\gamma2L_k) \Big).
\end{align*}

{\bf Step 7:} End of the proof.\\
For compact $K$, we find $k\ge k_1$ such that $K\subset\Lambda^{{\rm int}}_{L_k}(0)$. Then with $D = D(d, k, p, \| \eta \|_\infty)$, we have

\begin{align*}
\EE \Big\{ \| |X|^p\eta(H(\omega))\chi_K\| \Big\} \le &\,  c_d L_k^p \| \eta \|_\infty + \EE \Big\{ \sum_{j\ge k} \||X|^p \chi_{\Lambda_{L_{j+1}} \setminus \Lambda_{L_j}} \eta(H(\omega)) \chi_K\| \Big\}\\
\le &\, D + \sum_{j\ge k} c_d L_{j+1}^p  \sum_{ \substack{\tilde x \in \Lambda_{L_{j+1}} \setminus \Lambda_{L_j} \\ \tilde x \in \Gamma_j}} \EE \{ \| \chi^{{\rm int}}_{L_j,\tilde x}\eta(H(\omega)) \chi^{{\rm int}}_{L_j,0}\| \} \\
\le &\, D \Big[ 1 + \sum_{j\ge1} L_j^{\alpha p} L_j^{d(\alpha-1)} \Big( L_j^{2d(\alpha-1)-2\xi} + \exp(-\tfrac\gamma2L_j) \Big) \Big]\\
< &\, \infty,
\end{align*}
since $\alpha p+3d(\alpha-1)-2\xi<0$ and the $L_j$ grow fast enough.
\end{proof}

Although we cannot apply the theorem directly, a look at the proof, 
particularly at Steps 5 to 7, shows that we have the following:

\begin{corollary}\label{c32}Let the assumptions of Theorem {\rm
    \ref{t31}} be satisfied. Then we have
$$ 
\EE\Big\{\sup_{t>0}\|
|X|^pe^{-itH(\omega)}P_I(H(\omega))\chi_K\|\Big\}<\infty.
$$
\end{corollary}

\section{Applications}
In this section we present a list of models for which the variable
energy multi-scale analysis has been established and which therefore
exhibit strong dynamical localization by the results of the preceding
section.

\subsection{Periodic plus Anderson}
Here we discuss band edge localization for alloy-type models which
consist of a periodic background operator with impurities sitting on the 
periodicity lattice. We take $\ZZ^d$ as this lattice simply for notational
convenience; a reformulation for more general lattices presents
no difficulties whatsoever. Note that compared with most results
available in the literature, we assume minimal conditions on the
single-site measure:

\begin{enumerate}
\item Let $p=2$ if $d\le 3$ and $p>d/2$ if $d>3$.
\item Let $V_0\in L^p_{{\rm loc}}(\RR^d)$, $V_0$ periodic w.r.t. $\ZZ^d$ and
  $H_0=-\Delta+V_0$. 
\item Let $f\in L^p(\Lambda_1(0))$, $f\ge 0$ and $f\ge \sigma$ on
  $\Lambda_s(0)$ for some $\sigma>0$, $s>0$; $f$ is called the {\it single-site potential}.
\item Let $\mu$ be a probability measure on $\RR$, with $\supp\mu
  =[q_-,q_+]$, where $q_-<q_+\in\RR$; $\mu$ is called the {\it single-site
  measure}.
\item Let 
$$\Omega=[q_-,q_+]^{\ZZ^d},\PP=\bigotimes_{\ZZ^d}\mu \mbox{  on
  }\Omega$$
and $q_k:\Omega\to\RR$, $q_k(\omega)=\omega_k$.
\item Let 
$$
V_\omega(x):=\sum_{k\in\ZZ^d}q_k(\omega)f(x-k)
$$
and 
$$
H^{\rm A}(\omega)=-\Delta+V_0+V_\omega .
$$
\end{enumerate}

For an elementary discussion of this model and all the ingredients
necessary to prove localization, we refer to \cite{book1}; see also
\cite{k2s,ks2}. Note that to conform with standard notation, we denote by $p$ both the power of the moment operator in the dynamical bounds and the power defining the appropriate $L^p$ space the potentials have to belong to. This, however, should not lead to any real confusion.  

\begin{theorem}\label{t41}
Let $H^{\rm A}(\omega)$ be as above. Assume that the single-site measure $\mu$ is H\"{o}lder continuous, that is, there exists $a>0$ such that for every interval $J$ of length small enough, $\mu(J)\le |J|^a$. Denote $\Sigma = \sigma(H^{\rm A}(\omega))$ a.e.~and $E_0 = \inf \Sigma$. Let $p>0$. Then there exists $\epsilon_0>0$ such that for $\eta \in L^\infty(\RR)$ with $\supp \eta \subset [E_0,E_0+\epsilon_0]$ and compact $K$, we have
$$
\EE\{\| |X|^p\eta(H^{\rm A}(\omega))\chi_K\|\}<\infty .
$$
Moreover, for $I\subset [E_0,E_0+\epsilon_0]$ and $K$ compact:
$$
\EE \Big\{\sup_t\| |X|^pe^{-iH^{\rm A}(\omega)t}P_I(H^{\rm A}(\omega))\chi_K\| \Big\} < \infty .
$$
\end{theorem}
\begin{proof}[{\bf Proof.}]
It is well known that (INDY), (WEYL), (GRI) and (EDI) are satisfied if
we take for $H^{\rm A}_\Lambda$ the operator $H^{\rm A}$ restricted to $\Lambda$ with periodic boundary conditions. Due to \cite{k2s,fat} we have a
Wegner estimate of the form
$$
\PP \{ \dist( \sigma(H^{{\rm A}}_\Lambda(\omega)),E_0 ) \le \exp(-L^\Theta) \} \le C \cdot L^2 \cdot d \cdot \exp(-a L^\Theta),
$$
where $L$ denotes the sidelength of the cube $\Lambda$. In particular, $W(I_0,L,\Theta,q)$ is satisfied for a neighborhood $I_0$ of $E_0$, arbitrarily given $\Theta$ and $q$, and $L$ large enough. For given $p>0$, we can start the multi-scale induction with $2\xi>p$ by Lifshitz asymptotics.
\end{proof} 

Note that the above theorem includes the case of single-site potentials
with small support. Moreover, using Klopp's analysis of internal Lifshitz tails \cite{kl}, Veselic establishes the necessary initial length scale estimates at lower band edges in the case where $H_0$ exhibits a 
non-degenerate behavior at the corresponding edge \cite{v}, so the result above extends to this case. If one does not know that $H_0$ has a non-degenerate band edge, one can still derive an initial length scale estimate by requiring a disorder assumption. This, however, might put some restriction on the power $p$.

\begin{theorem} 
\label{t42}
Let $H^{\rm A}(\omega)$ be as above. Assume
\begin{itemize}
\item[{\rm (i)}] The single-site measure $\mu$ is H\"{o}lder continuous.
\item[{\rm (ii)}] There exists $\tau >d$ such that for small $h>0$,
$$
\mu([q_-,q_-+h])\le h^\tau\mbox{   and   }\mu([q_+-h,q_+])\le h^\tau .
$$  
\end{itemize}
Denote $\Sigma=\sigma(H(\omega))$ a.e.~and let $E_0 \in \partial \Sigma$.
Let $p<2(2\tau-d)$. Then there exists $\epsilon_0>0$ such that for $\eta \in L^\infty(\RR)$ with $\supp \eta \subset [E_0 - \epsilon_0, E_0 + \epsilon_0]$ and compact $K$, we have
$$
\EE \{\| |X|^p\eta(H^{\rm A}(\omega))\chi_K\|\}<\infty .
$$
Moreover, for $I\subset [E_0-\epsilon_0,E_0+\epsilon_0]$ and $K$ compact:
$$
\EE \Big\{\sup_t\| |X|^pe^{-iH^{\rm A}(\omega)t}P_I(H^{\rm A}(\omega))\chi_K\| \Big\} < \infty .
$$
\end{theorem}
\begin{proof}[{\bf Proof.}] We have already checked everything except for the initial length scale estimate $G(I,L,\gamma,\xi)$, and in particular how large $\xi$ can be taken. By an elementary argument, we can take $\xi$ subject to the condition $\xi<2\tau-d$ (see \cite{k2s}), which gives the claimed result.
\end{proof}

With a modification of independent multi-scale analysis given in
\cite{ks2} we can also treat the correlated or long-range case, by which 
we understand that the single-site potential $f$ is no longer assumed
to have support in the unit cube; see \cite{ks2}.

\begin{theorem} 
\label{t43}
Let $H^{\rm A}(\omega)$ be as above, with condition {\rm 3} replaced by
\begin{itemize}
\item[{\rm 3.}] Let $f\in L^p_{{\rm loc}}$, $f\ge 0$ and $f\ge \sigma$ on
  $\Lambda_s(0)$ for some $\sigma>0$, $s>0$; 
$$
f\le C |x|^{-m}\mbox{ for }|x| \mbox{ large.}
$$ 
\end{itemize}
Then the conclusions of {\rm Theorems \ref{t41}} and {\rm \ref{t42}} hold true with 
$$
p<2\left(\frac{m}{4}-d\right)\mbox{   and   }p<2\left(\frac{m}{4}-d\right)\wedge 2(2\tau-d),
$$
respectively.
\end{theorem}

{\it Remark.} Although discrete models are not explicitly included in the above framework, our principal strategy pursued in Section 3 is clearly able to treat random operators in $\ell^2(\ZZ^d)$ for which a multi-scale analysis has been established. In particular, building on results from \cite{ckm} one may establish strong dynamical localization for the discrete Anderson model, where for $d=1$, even pure point single-site measures (e.g., the Bernoulli case) are within the scope of this result. See \cite{ckm} for explicit requirements to make the multi-scale machinery work. We thus obtain new results on strong dynamical localization also in the discrete case since the Aizenman method does not cover single-site distributions which are too singular (e.g., the Bernoulli case).

\subsection{Random divergence form operators}
The following type of model has been introduced in \cite{fk,acoustic}
in order to study classical waves (see \cite{fk} for a motivation). These 
models are also intensively studied in \cite{book1}.
\begin{enumerate}
\item Let $\aa_0:\RR^d\to M(d\times d)$ be measurable, $\ZZ^d$-periodic and such that for some $\eta>0$, $M>0$,
$$
\eta\le\aa_0(x)\le M \mbox{   for all   }x\in \RR^d
$$
as matrices, that is, $\eta\|\zeta\|^2\le (\aa_0(x)\zeta|\zeta)\le M \| \zeta \|^2$ for every $\zeta\in\CC^d$.
\item Let $S=[0,\lambda_{\max} ]^d \times \mathcal{O}(d)$, where $\lambda_{\max} > 0$ and $\mathcal{O}(d)$ denotes the orthogonal matrices. 
\item Let $\nu$ be a probability measure on $\mathcal{O}(d)$ and let $\gamma_i$, $i=1,...,d$ be probability measures on $\RR$ with $\supp\gamma_i=[0,\lambda_{\max} ]$. 
\item $S$ is called the {\it single-site space} and $\mu = \gamma_1 \otimes
\cdots \otimes\gamma_d\otimes\nu$ is called the {\it single-site measure}.
\item Let 
$$
\Omega=S^{\ZZ^d}, \,\, \PP=\mu^{\ZZ^d},
$$
and for $\omega(k)=(\lambda_1(k),...,\lambda_d(k),u(k))$, define
$$\aa_k(\omega)=u(k)^*\diag(\lambda_1(k),...,\lambda_d(k))u(k),$$
where
$\diag(\lambda_1(k),...,\lambda_d(k))$ denotes the diagonal matrix with
the indicated diagonal elements. 
\item Define
$$\ao(x):=\sum_{k\in\ZZ^d}\chi_{\Lambda_1(k)}(x)\aa_k(\omega)$$
and 
$$ H^{{\rm DIV}}(\omega)=-\nabla(\aa_0+\ao)\nabla .$$
\end{enumerate}

Although the formulas may seem intricate, it is easy to see what is
happening. For site $k$, we choose a non-negative matrix $\aa_k(\omega)$
at random by choosing its $d$ eigenvalues and a unitary conjugation matrix. This is done independently at different sites and we get an Anderson-like
random matrix function $\ao$ which is used as a perturbation to the
perfectly periodic medium $\aa_0$. Note that $\aa_0+\ao$ have uniform
upper and lower bounds ($\eta$ and $M+\lambda_{\max} $) so that the operators can be defined via quadratic forms with the Sobolev space $W^{1,2}(\RR^d)$ as common form domain. The initial value problem we are now interested in is governed by the wave equation
$$
\frac{\partial^2 v}{\partial t^2} = -H^{{\rm DIV}} (\omega)v, \quad v(0) = v_0, \quad \frac{\partial v}{\partial t}|_{t=0} = v_1 \eqno{{\rm (WE)}}
$$
rather than the Schr\"odinger equation. Solutions are given by
$$
v(t)=\cos \Big( t \sqrt{H^{{\rm DIV}}(\omega)} \Big) v_0 + \sin \Big( t \sqrt{H^{{\rm DIV}}(\omega)} \Big) w_1,
$$
where $v_1=\sqrt{H^{{\rm DIV}}(\omega)}w_1$, and $v_0,w_1$ have to belong to the appropriate operator domains. The following result yields a strong form of dynamical localization in this case:

\begin{theorem}\label{t44}
Let $H^{{\rm DIV}}(\omega)$ be as above. Assume
\begin{itemize}
\item[{\rm (i)}] The measures $\gamma_i$, $i=1,...,d$ are H\"{o}lder continuous.
\item[{\rm (ii)}] There exists $\tau >d$ such that for small $h>0$,
$$
\gamma_i([0,h])\le h^\tau\mbox{   and   }\gamma_i([\lambda_{\max} -h,\lambda_{\max} ])\le h^\tau  
$$
for all $ i=1,...,d$.
\end{itemize}
Denote $\Sigma=\sigma(H^{{\rm DIV}}(\omega))$ a.e.~and let $E_0 \in \partial \Sigma \setminus \{ 0\}$.

Then there exists $\epsilon_0>0$ such that for $\eta \in L^\infty(\RR)$ with $\supp \eta \subset [E_0-\epsilon_0,E_0+\epsilon_0]$ and compact $K$, we have
$$
\EE \{\| |X|^p\eta(H^{{\rm DIV}}(\omega))\chi_K\| \} < \infty .
$$
Moreover, for $I\subset [E_0-\epsilon_0,E_0+\epsilon_0]$ and $K$ compact:
$$
\EE \Big\{ \sup_t\| |X|^p \cos \left( t \sqrt{H^{{\rm DIV}}(\omega)} \right) P_I (H^{{\rm DIV}}(\omega))\chi_K\| \Big\} < \infty
$$
and
$$
\EE \Big\{ \sup_t\| |X|^p \sin \left( t \sqrt{H^{{\rm DIV}}(\omega)} \right) P_I (H^{{\rm DIV}}(\omega))\chi_K\| \Big\} < \infty .
$$
\end{theorem}
By the results from \cite{fk,acoustic} the conditions for multi-scale
analysis are satisfied.

\subsection{Random quantum waveguides}
Quantum waveguides have been introduced for the investigation
of two- or three-dimensional motion of electrons in small channels, tubes
or layers of crystalline matter of high purity. Mathematically speaking, 
one considers the free Laplacian in a domain which should be thought of
as a perturbation of a strip. The following random model is taken from
\cite{klst}, where all the necessary conditions for multi-scale
analysis are verified:

It consists of a collection of randomly dented versions of a parallel
strip $\mathbb{R} \times (0,d_{\max}) = D_{\max}$. More precisely, let
$d_{\max}>0$, $0 < d < d_{\max}$, and consider $\Omega = [0,d]^\mathbb{Z}$. The $i$-th coordinate $\omega(i)$ of $\omega \in \Omega$ gives the deviation
of the width of the random strip from $d_{\max}$, that is,
\[ d_i(\omega) := d_{\max}-\omega(i),\]
which lies between $d_{\min}=d_{\max}-d$ and $d_{\max}$. Define
$\gamma(\omega): \mathbb{R} \to [d_{\min},d_{\max}]$ as the polygon in $\mathbb{R}^2$
joining the points $\{(i,d_i(\omega))\}_{i\in \mathbb{Z}}$ and 
\[ D(\omega) = \{(x_1,x_2)\in \mathbb{R}^2 \,|\, 0 < x_2 <
\gamma(\omega)(x_1)\}. \]
The following picture will help in visualizing this domain:

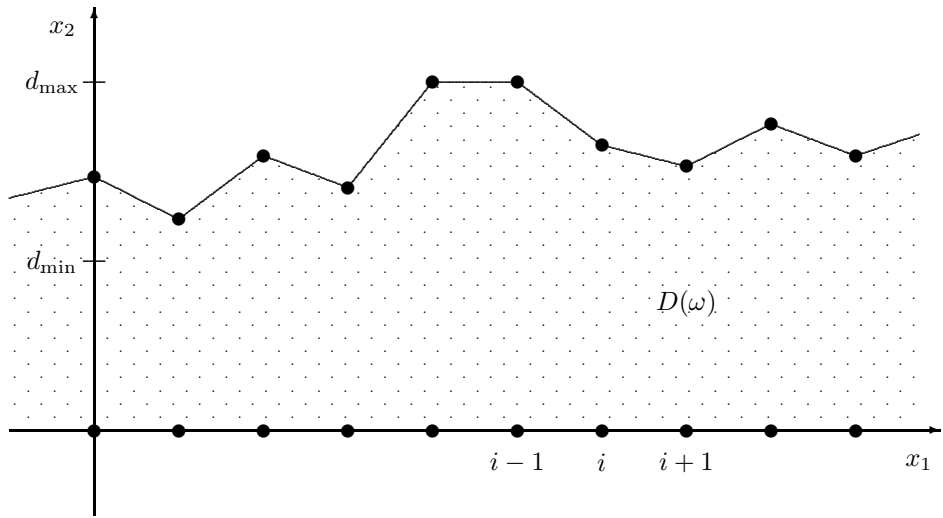
\[
\beginpicture
\setcoordinatesystem units <0.8pt,0.8pt> point at 40 40 
\put{\vector(1,0){352}} [Bl] at -40 0
\put {$x_1$} at 390 -15 
\put{\vector(0,1){192}} [Bl] at 0 -40 
\put {$x_2$} at -15 190
\plot -5 80 5 80 /
\put {$d_{\min}$} at -20 80
\plot -5 165 5 165 /
\put {$d_{\max}$} at -20 165
\multiput{\circle*{5}} [Bl] at 0 0 *9 40 0 /
\put {$i-1$} at 200 -15
\put {$i$} at 240 -15
\put {$i+1$} at 280 -15 
\put{\circle*{5}} [Bl] at 0 120
\put{\circle*{5}} [Bl] at 40 100
\put{\circle*{5}} [Bl] at 80 130 
\put{\circle*{5}} [Bl] at 120 115
\put{\circle*{5}} [Bl] at 160 165
\put{\circle*{5}} [Bl] at 200 165
\put{\circle*{5}} [Bl] at 240 135
\put{\circle*{5}} [Bl] at 280 125
\put{\circle*{5}} [Bl] at 320 145
\put{\circle*{5}} [Bl] at 360 130
\setlinear
\plot -40 110 0 120 40 100 80 130 120 115 160 165 200 165 240 135 280
125 320 145 360 130 390 140 /
\vshade -40 0 110 <,z,,> 0 0 120 <z,z,,> 40 0 100 <z,z,,> 80 0 130
<z,z,,> 120 0 115 <z,z,,> 160 0 165 <z,z,,> 200 0 165 <z,z,,> 240 0 135
<z,z,,> 280 0 125 <z,z,,> 320 0 145 <z,z,,> 360 0 130 <z,,,> 390 0 140 /
\put {$D(\omega)$} at 280 60
\endpicture
\]

We fix a probability measure $\mu$ on $[0,d]$ with $0 \in \supp \mu \neq \{0\}$ and introduce $\mathbb{P} = \mu^\mathbb{Z}$, a probability measure on $\Omega$. Consider $H^{\rm W}(\omega) = - \Delta_{D(\omega)}$, the Laplacian on $D(\omega)$ with Dirichlet boundary conditions, which is a self-adjoint operator in $L^2(D(\omega))$. 
\medskip
Note that
$$
\inf\sigma(H^{\rm W}(\omega))=E_0:=\frac{\pi^2}{d_{\max}^2} \mbox{ for }
\mathbb{P}\mbox{-a.e. } \omega \in \Omega.
$$
In \cite{klst}, exponential localization in a neighborhood of $E_0$ is
proven with the help of a variable energy multi-scale analysis similar
to the one presented in Section 2 of the present paper. In particluar, suitably modified versions of the assumptions of Theorem \ref{t31} are established which enable one to prove strong dynamical localization along the lines of Section 3. Thus, we have

\begin{theorem}\label{t45}
Let $H^{\rm W}(\omega)$ and $E_0$ be as above. Assume that the single-site measure $\mu$ is H\"{o}lder continuous. Let $p>0$. Then there exists $\epsilon_0>0$ such that for $\eta \in L^\infty (\RR)$ with $\supp \eta \subset [E_0, E_0 + \epsilon_0]$ and compact $K$, we have
$$
\EE\{\| |X|^p\eta(H^{\rm W}(\omega))\chi_K\|\}<\infty .
$$
Moreover, for $I\subset [E_0,E_0+\epsilon_0]$ and $K$ compact:
$$
\EE \Big\{ \sup_t \| |X|^pe^{-iH^{\rm W}(\omega)t}P_I(H^{\rm W}(\omega))\chi_K\| \Big\} < \infty .
$$
\end{theorem}
\subsection{Landau Hamiltonians}
The models we discuss now are particularly interesting due to their
importance for the quantum Hall effect and hence have been studied
intensively \cite{ch, dmp1, dmp2, gdb, w}. We rely here on the set-up
from \cite{ch}, also considered in \cite{gdb}, as the latter authors
provide a proof of the basic assumptions needed for our approach. In
particular, the trace condition (i) from Theorem \ref{t31} is proven
there and the validity of (GRI) is discussed.

We consider electrons confined to the plane $\RR^2$ subject to a
perpendicular constant $B$-field.

Assume
\begin{enumerate}
\item $H_0=\left( \partial_1+\frac{B}{2}x_2\right)^2+ \left( \partial_2 - \frac{B}{2} x_1 \right)^2$, where $B>0$ is constant.
\item Let $\supp f\in L^\infty(\Lambda_1(0))$, $f\ge 0$ and $f \ge \sigma$ on $\Lambda_s(0)$ for some $\sigma>0$, $s>0$; $f$ is called the 
{\it single-site potential}.
\item Let $\mu$ be a probability measure on $\RR$, with density $g$,
$g \in C^2_0(\RR)$ even and strictly positive a.e.~on its support $[-q,q]$; $\mu$ is called the {\it single-site measure}.
\item Let
$$
\Omega=[-q,q]^{\ZZ^2},\, \PP=\bigotimes_{\ZZ^2}\mu \mbox{ on }\Omega
$$
and $q_k:\Omega\to\RR$, $q_k(\omega)=\omega_k$.
\item Let
$$
V_\omega(x):=\sum_{k\in\ZZ^2}q_k(\omega)f(x-k)
$$
and
$$
H^{\rm L}(\omega)=H_0+V_\omega .
$$
\end{enumerate}
Recall that the spectrum of $H_0$ in this case consists of the sequence
of Landau levels $E_n(B)=(2n+1)B$. We have the following:

\begin{theorem}\label{t46}
Let $H^{\rm L}(\omega)$ be as above with $B$ large enough. Let $p>0$. Then for every $n \in \NN$, there exists $\epsilon_n(B) = O(B^{-1})>0$ such that for $\eta \in L^\infty(\RR)$ with $\supp \eta \subset [E_n(B) + \epsilon_n(B), E_{n+1}(B) - \epsilon_n(B)]$ and compact $K$, we have
$$
\EE\{\| |X|^p\eta(H^{\rm L}(\omega))\chi_K\|\}<\infty .
$$
Moreover, for $I \subset [E_n(B) + \epsilon_n(B),E_{n+1}(B) - \epsilon_n(B)]$ and $K$ compact:
$$
\EE \Big\{ \sup_t\| |X|^pe^{-iH^{\rm L}(\omega)t}P_I(H^{\rm L}(\omega)) \chi_K\|\Big\} < \infty .
$$
\end{theorem}

In \cite{w}, a proof of exponential localization is given for a case
which includes single-site potentials of changing sign. However, the
use of microlocal techniques requires smoothness of the potential.
\\[5mm]
{\it Acknowledgements.} The results presented here were obtained
during a visit of P.~S.~to Caltech. Partial financial support by the DFG
and the hospitality at Caltech and Santa Monica are gratefully
acknowledged. D.~D.~received financial support from the German Academic Exchange Service through Hochschulsonderprogramm III (Postdoktoranden). Moreover, we would like to thank F.~Germinet, S.~Jitomirskaya and A.~Klein for helpful discussions.

\end{document}